\documentclass{elsarticle}
\usepackage{amsmath}
\usepackage{latexsym}
\journal{Journal of Nuclear Materials}
\begin{document}
\begin{frontmatter}

\title{Strike point splitting induced by the application of magnetic perturbations on MAST}
\author[ipp]{P. Cahyna\corref{cor1}\fnref{pres}}
\ead{cahyna@ipp.cas.cz}
\author[ipp]{M. Peterka}
\author[ccfe]{A. Kirk}
\author[ccfe]{A. Thornton}
\author[ccfe]{J. Harrison}
\author[ccfe]{D. Muir}
\author[ipp]{R. Panek}
\author{the MAST team}
\address[ipp]{Institute of Plasma Physics, AS CR, v.v.i., Association EURATOM/IPP.CR, Za Slovankou 3, 182 00 Prague, Czech Republic}
\address[ccfe]{EURATOM/CCFE Fusion Association, Culham Science Centre, Abingdon, Oxon, OX14 3DB, UK}
\fntext[pres]{Presenting author}
\cortext[cor1]{Corresponding author}

\begin{abstract}
  Divertor strike point splitting induced by resonant magnetic
  perturbations (RMPs) has been observed on MAST for a variety of RMP
  configurations in a plasma scenario with $I_p = 750\;\text{kA}$
  where those configurations all have similar resonant
  components. Complementary measurements have been obtained with
  divertor Langmuir probes and an infrared camera. Clear splitting
  consistently appears in this scenario only in the even configuration
  of the perturbation coils, similarly to the density pump-out. These
  results present a challenge for models of plasma response to RMPs.
\end{abstract}


\end{frontmatter}

\newpage 
\section{Introduction}
The application of resonant magnetic perturbations (RMPs) to tokamak
plasma with an X-point is expected to result in formation of a 3-D
structure of magnetic field lines --- the homoclinic tangle, replacing
the original separatrix~\cite{Evans05}. This should in turn result in
formation of spiralling structures of fluxes (footprints) on the
divertor as parallel transport carries particles and energy along the
field lines inside the tangle from the core plasma directly to the
divertor plates. Such perturbations are used for ELM control and
foreseen for this use on ITER~\cite{thomasiaea2008}, and during the
ELM control experiments such structures are indeed observed. The
observation is usually performed on a single toroidal location on the
divertor along a radial profile. The strike point is then observed to
split into multiple peaks which correspond to the intersections of the
diagnostic line of sight with the footprints. The experimental evidence for
the formation of footprints has been mixed though. Generally the
effect is more consistently observed in L-mode than in H-mode
\cite{NardonPSI2010,SchmitzPSI2010}, and the observations also depend
on the quantity being measured: heat flux, measured by the infrared
(IR) camera, shows much less clear splitting than visible light
emission in DIII-D low-collisionality H-mode discharges
\cite{SchmitzPSI2010} (results are again different for high
collisionality~\cite{0029-5515-49-9-095013}).

From the theoretical point of view the interesting feature of divertor
footprints is that they are determined by the same quantities as the
resonant perturbation modes which determine the width of magnetic
islands and thus the transition to stochastic regime of field lines
--- the Melnikov \cite{joseph2008,pavel-eric-footprints} or Poincar\'e
\cite{abdullaev:042508} integrals. One of the open questions of the
RMP technique is to what extent the plasma screens the resonant modes
and thus reduces the edge stochastic layer to much lesser width than
predicted by vacuum calculations. Such screening is predicted by
extended MHD \cite{ericmodel} and kinetic \cite{heyn2008} models
especially for H-mode due to strong gradients and flows in the
pedestal. Footprints are then predicted to be significantly reduced in
comparison to results based on vacuum field \cite{pavel-eric-psi} when
screening is introduced in the model. The observations of footprints
thus provide a valuable test for screening models --- the differences
in experimental results may be partly due
to differences in plasma response
\cite{NardonPSI2010}.

Significant number of experiments with RMPs have been performed on the
MAST tokamak~\cite{ericeps09,kirksfp09,0741-3335-53-6-065011} since
the installation of the in-vessel RMP coils~\cite{ericpsi}. One
notable effect of RMP application on MAST is the density pump-out in
L-mode. It is remarkable that in a scenario with 750~kA of plasma
current the pitch angle of edge field lines is such that the odd and
even configuration of the perturbation field have similar resonant
components. This results in an almost identical Chirikov factor
profile, however the density pump-out has been observed in the even
configuration only~\cite{0741-3335-53-6-065011}. The
pump-out thus can not be explained by any mechanism relying purely on the
formation of a stochastic layer in the vacuum field. Even introduction
of screening can hardly explain the difference, as the screening
results from an interplay between the resonant modes, plasma flows and
other plasma parameters (resistivity, viscosity), which are here same
or similar for the even and odd configurations in the same plasma
scenario. One possibility proposed in~\cite{0741-3335-53-6-065011} is
that the plasma response has a different form for even and odd
perturbation, as measured by the displacement calculated by the MHD
model MARS-F. In any case, thanks to the unexpected results, this
scenario represents an important testbed of plasma response models. At
the same time the fact that it is in L-mode simplifies modelling since
it does not need to take into account the strong gradients and flows
in H-mode.

Due to the importance of the aforementioned experiments for the
understanding of plasma response to RMPs and the ability of divertor
footprints to reveal details of the magnetic field we present in this
paper an analysis of the strike point data available for these
shots. As in the previous reports on divertor footprint observations on
MAST~\cite{NardonPSI2010,ericeps09} we used an IR camera to measure
temperature and infer heat flux to the divertor targets. In addition
we use here for the first time on MAST the divertor Langmuir probe
measurements to enlarge the base of available data, following the
example of DIII-D~\cite{Watkins2009839}.

\section{Experimental setup and methods}
\label{sec:setup-methods}
The plasma scenario used in all the shots presented here is a
double-null, L-mode plasma with a current of 750~kA. The set of
perturbation coils was described in~\cite{ericpsi}. The coils are
wired in an $n=3$ configuration, either even or odd parity. For each
parity there are two possible phases, labeled as $0^\circ$ and
$60^\circ$, yielding four possible configurations in total. In all the
shots studied here the current in the  RMP coils had the same time evolution.
During a shot the strike point position is evolving due to the influence of
the changing field of the central solenoid. The data collected from
the divertor at a
given toroidal position are a function of two variables: time and
radial position on the divertor plate. If we plot this function we see
the motion of the strike point and features associated with it, such
as the secondary maxima due to strike point splitting. In contrast
there can be patterns due to irregularities in the divertor surface,
or in the case of probes, to one probe having a systematic error. Those patterns are static and thus the
motion of strike points allows us to distinguish them from the real
strike point splitting.

The strike
point on the lower divertor was observed by an IR camera with a
wavelength range of 4.5 to 5.0~$\mu\text{m}$ and temporal resolution of
1250~$\mu\text{s}$. The camera was zoomed on a radial band on the divertor
with a size of $320\times 96$ pixels, the spatial resolution was
1~mm. From the measured temperature of the divertor plate the incoming
heat flux is estimated.

In addition there are six arrays of Langmuir probes in the divertor
plates, three in the lower and three in the upper divertor at
toroidal locations with toroidal angles of $\varphi=233^\circ,
288^\circ, 333^\circ$. Distance
between consecutive probes is 9.23~mm. The probes are being
continually swept and the I-V characteristics is measured, which is
then fitted in order to obtain the floating potential
$V_{\text{float}}$, ion saturation current $J_{\text{sat}}$, electron
temperature $T_e$ and an estimate of the incoming power. Each voltage sweep takes 65~$\mu \text{s}$ but due
to multiplexing each probe is only swept once every 1.04~ms. The spatial resolution of the probes is rather low for
measuring structures such as the divertor footprints. The resolution
can be significantly enhanced by using the strike point motion which
causes a given probe to measure points with varying distance from the
primary strike point as the whole footprint structure is moving across
it. To exploit this we divide the divertor space in bins with fixed
width and distance from the primary strike point whose position is
determined by equilibrium reconstruction (using the EFIT code) and in
every bin we perform an averaging over the values which fall into it
during the time evolution. We have been using a bin width of
6.5~mm. It would be possible to use even smaller bins to further
improve the radial resolution, but due to fairly significant noise in
the data choosing larger bins produces data of better quality thanks
to averaging of more values. This technique requires the relative
position of the EFIT prediction and the actual strike point to be
unchanging during a shot, although an agreement on the absolute
position is not necessary.

\begin{figure}[htb]%
  \centering%
  \includegraphics[width=160mm]{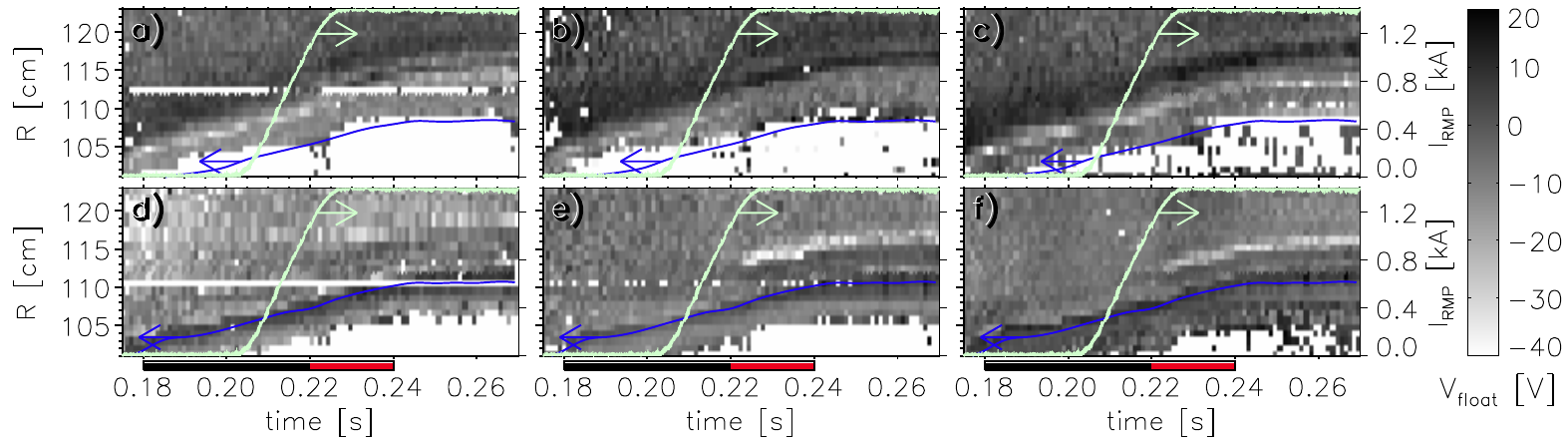}%
  \caption{Time dependence of $V_{\text{float}}$ on all the sectors of
  divertor probes during the shot 25941. a) -- c) upper sectors, d) --
  f) lower
    sectors. Blue line: position of the
  strike point as calculated by EFIT ($R_\text{EFIT}$). Cyan line: current in the
  perturbation coils $I_\text{RMP}$, with a maximum of 1.4 kA. Black
  and red bands below the graph indicate the time windows used for
  averaging in the following figures.}
\label{fig:conf9map}
\end{figure}

This method on the other hand significantly reduces the time
resolution. We have been using only two time windows for averaging,
giving us two time points. The first window is for times from 0.18~s
to 0.22~s and the second one from 0.22~s to 0.24~s (the windows are
shown as color bars in Fig.~\ref{fig:conf9map}). The boundary
(0.22~s) was chosen as at this time structures usually start to
appear (see Fig.~\ref{fig:conf9map}) and the current in the
RMP coils nearly reaches its maximum value (\emph{ibid.}).

\section{Results}
\begin{figure}[htb]%
  \centering%
  \includegraphics[width=75mm]{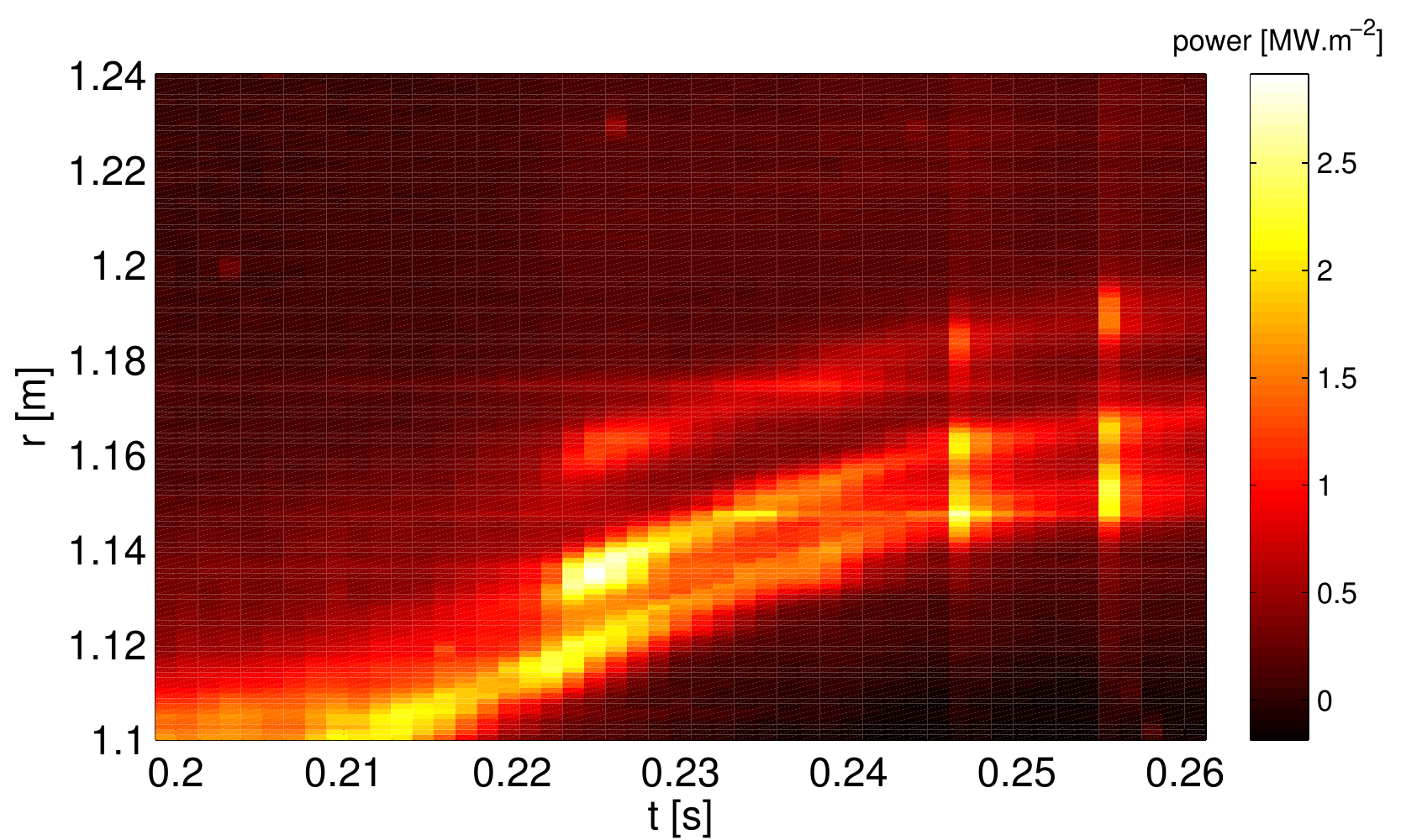}%
  \caption{Time dependence of the heat flux profile to the divertor
    target measured by the IR camera for shot 25965.}
\label{fig:2dcamera}
\end{figure}

\begin{figure}[htb]%
  \centering%
  \includegraphics[width=75mm]{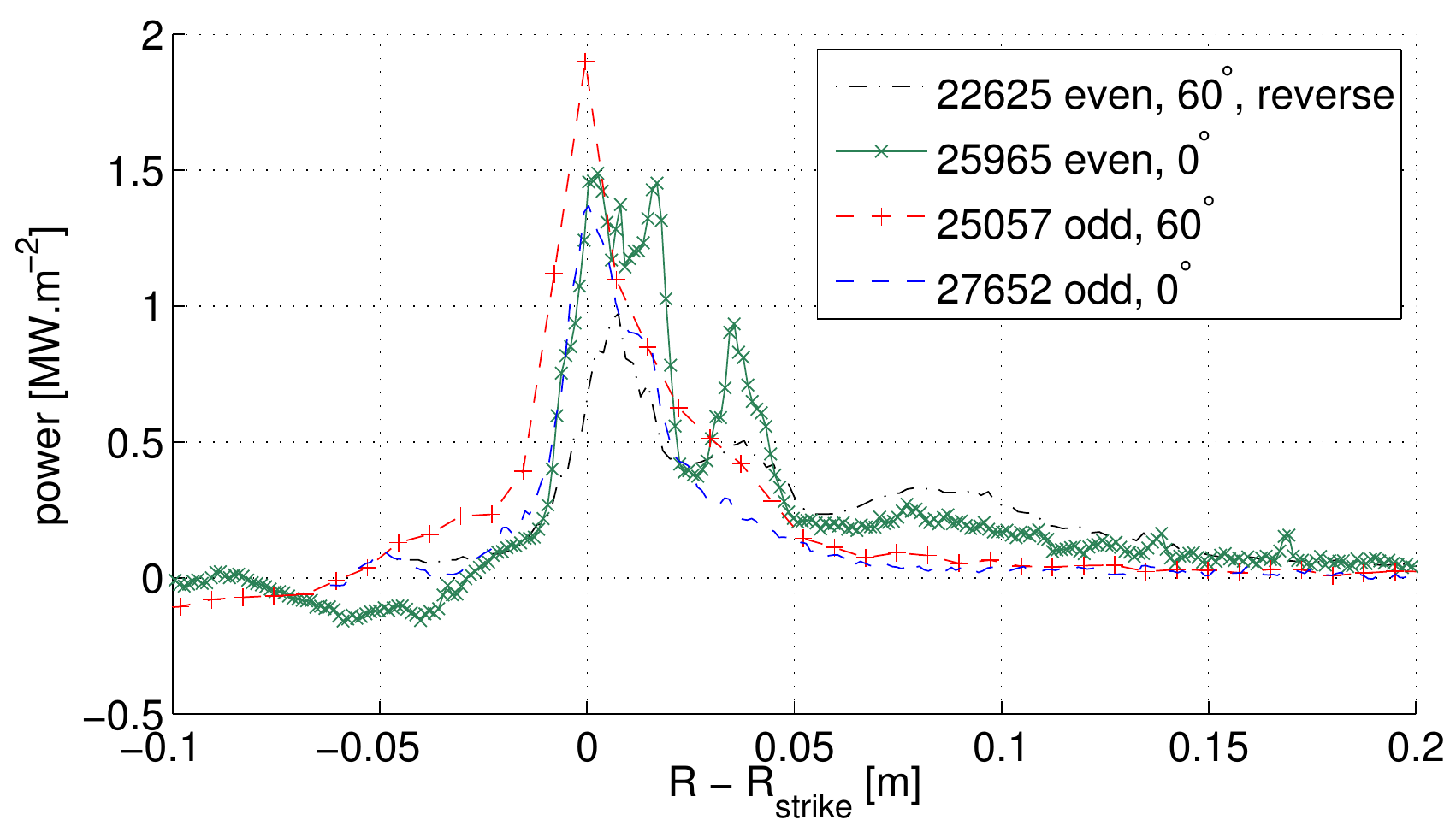}%
  \caption{Heat flux profiles to the divertor target measured by the
    IR camera as a function of distance from the primary strike point
    for several shots at $I_\text{RMP}=1.4\;\text{kA}$ at time
    $t=0.24\;\text{s}$.}
\label{fig:1dcamera}
\end{figure}

The list of shots discussed below and shown in the figures is given in
table~\ref{tab1}. 
The IR camera observes in many cases a clearly recognizable splitting
of the strike point after the perturbation has been switched on. In
some cases however there is no  or almost undistinguishable splitting. In other cases there is
a clearly visible peak about 4~cm away from the strike line
(Fig.~\ref{fig:2dcamera}). Its position agrees with the position of
the magnetic footprint predicted by vacuum modelling. The
analysis of the available shots reveals that the even parity $60^\circ$
phase shots consistently exhibit the former result while the even
parity $0^\circ$ phase shots the latter. One exception is the shot
22625 which had opposite current and toroidal field than the other shots
and in which the perturbation was of an even parity $60^\circ$
phase. Here the result is similar to the shots with $0^\circ$ phase
(splitting with a clearly defined secondary peak at about 4~cm outside
the primary one, Fig.~\ref{fig:1dcamera}). There had been only few shots in
odd parity. The shots with $60^\circ$ phase did not show any
splitting (Fig.~\ref{fig:1dcamera}), however the camera was not zoomed so the resolution was
inferior to the other shots. The shot with $0^\circ$ phase had the same
camera setup (zoomed) as most of the other shots. Here splitting was
not observed either (Fig.~\ref{fig:1dcamera}).

The time evolution of the floating potential $V_{\text{float}}$ from
probes in all the available six sectors for the shot 25941 (even parity
$60^\circ$ phase) is shown in
Fig.~\ref{fig:conf9map} together with the evolution of strike points predicted by
EFIT and the time trace of the current in the perturbation coils $I_{\text{RMP}}$. The
predicted strike point follows the same path as the actual maxima and
minima of $V_{\text{float}}$, however we see that the absolute
position of predicted strike point is not correct and moreover the
displacement is not
the same in different sectors. The appearance of multiple bands when the
coils are switched on can be clearly seen, despite the lack of clear
splitting in IR
images for this shot. Note that the separation of
the bands is different in each sector. The profile of
$V_{\text{float}}$ is rather different in every sector even before the
coils are switched on and is not monotonic with increasing distance
from the strike points.

\begin{figure}[htb]%
  \centering%
  \includegraphics[width=160mm]{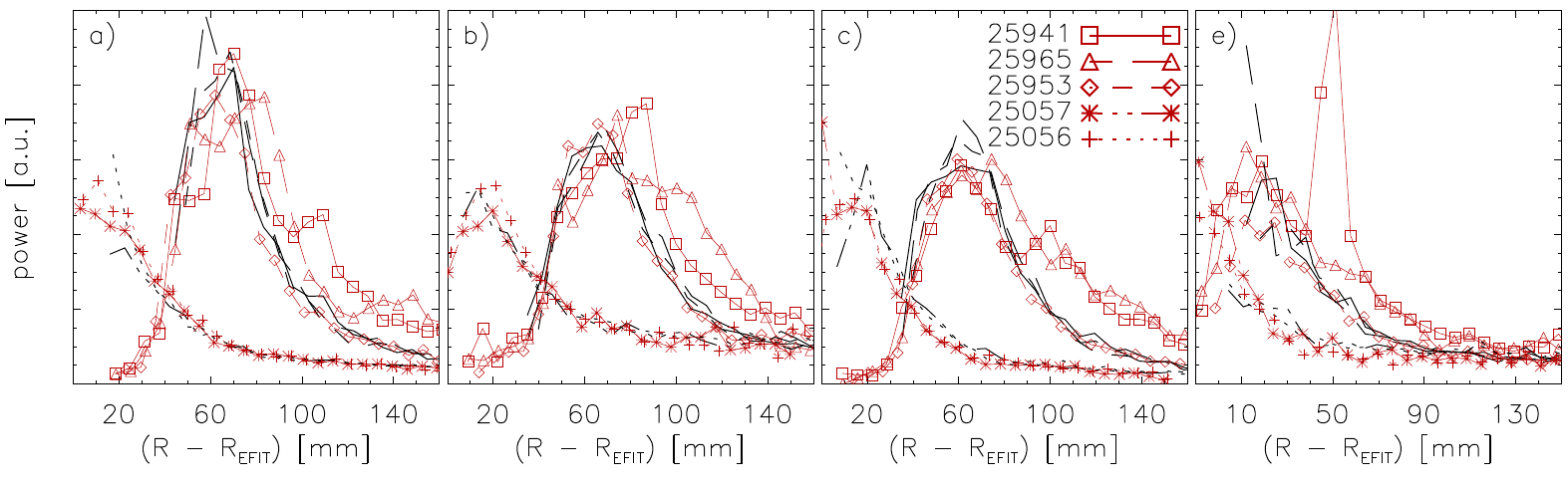}%
  \caption{Comparison of heat flux profiles measured by the divertor
    probes for various perturbation coil configurations. Sectors are
    labeled according to Fig.~\ref{fig:conf9map}. Black lines without
    symbols: time window before $t=0.22\;\text{s}$, red lines with
    symbols: time window after $t=0.22\;\text{s}$. Line styles
    distinguish shots and have the same meaning for black and red
    lines.}
\label{fig:moreprobes}
\end{figure}

Figure \ref{fig:moreprobes} shows the profiles of the power flux averaged over
two time windows before (black lines) and after (red lines) the coils
are switched on in four sectors. (In the remaining two the data are
very noisy and not usable.) Shots with maximum coil current in
different configurations and reference shots with zero
coil current are compared. The profiles are shown as a function of distance from
the strike point predicted by EFIT, as described above. The shots with
odd parity configuration are from another experimental campaign than
the others, here changes in both the EFIT prediction and the actual
strike points position result together in a large displacement of the
profile.

\section{Discussion}
IR camera and divertor probes can both reveal strike point splitting
and both have advantages and disadvantages. The IR camera has the
advantage of superior space resolution and less noise,
however it usually observes only one location and more observation
points, while desirable due to the 3D nature of the effect, are
difficult to achieve due to the significant price of the
equipment. For this reason it is advisable to have also divertor probe
arrays when feasible. We saw that the additional data from divertor
probes enabled us to confirm the existence of splitting in a scenario
(even $60^\circ$ phase) where the camera did not detect it. The
disadvantage of probes is the much inferior quality of their data. The
clearest picture of strike point splitting was obtained using the
$V_{\text{float}}$ measurement. It is remarkable that there is a
significant drop of the minimum of the floating potential on the lower
probe arrays, reminiscent of the DIII-D results~\cite{Watkins2009839} and
possibly indicating a presence of field lines connected to the core
plasma. The
floating potential observations are only qualitative
though, as the dependence of $V_{\text{float}}$ on the distance from
the strike point is complicated and moreover different for each probe
array. For this reason we have been using the power measurements,
which show a simple exponential decrease with the distance from the
strike points. This enables us to detect the effect of perturbation
coils even in cases where there are no clear secondary peaks but the
profile just broadens. The power measurements have much more noise
than the $V_{\text{float}}$ measurements, which are remarkably clean
even in the two sectors where the power (and $J_{\text{sat}}$, $T_e$)
data are too noisy to be useful. To eliminate the noise and systematic
errors  and increase the spatial resolution time averaging has been
useful. The resulting profiles show large fluctuations at the peak,
but the exponential decrease outside the strike point is
remarkably smooth and very precisely overlaps for different shots before the
coils are switched on and also in reference shots without
perturbation. We can thus claim that the averaged profiles have
nagligible error and the observed profile broadening in cases with
perturbation is a real effect. The probes have also the advantage of
being fixed, while the camera set-up can be changed or the camera can
be removed according to the needs of the experiments. In principle, if
all the auxiliary equipment  (power supplies, amplifiers) remains identical,
the probes can provide repeatable and comparable measurements.

The results show no clear splitting in odd configuration. To show
that this is a consequence of the actual divertor footprints being
small or absent it is necessary to compare several experimental
profiles and coil configurations. For the even parity, in the
$60^\circ$ phase no significant splitting was observed either on the IR
camera, in contrast with the $0^\circ$ phase. The explication may be
that the phase change rotates the footprints from the camera view. To
prove that this is the case (rather than footprints being absent) one
can use the probe data, which show actually more clear splitting in
the $60^\circ$ phase. This is supported by the fact that with an
opposite plasma current and toroidal field the footprints were clearly
seen on camera. It can be shown that a reversal of both the poloidal and
toroidal fields is for the footprint geometry equivalent to a reversal
of the perturbation phase. The position of peaks measured by probes is
different for both phases and for different toroidal sectors. The
latter difference is in agreement with the 3D nature of the
footprints. As the negative result for the odd
configuration was confirmed for both phases on both the camera and
probes, we can confidently claim that the footprints are absent or
significantly smaller than in the even configuration. In the future we
will investigate if this difference may be due to the same difference
in plasma response, as calculated by MARS-F, that was found to
correlate with the density pump-out. If this were not the case then
clearly the cause of the differences between odd and even parity
should be looked for elsewhere.

\section{Acknowledgements}
Discussions with E. Nardon are highly appreciated. This work,
part-funded by the European Communities under the contracts of
Association between EURATOM and IPP.CR and CCFE, was carried out
within the framework of the European Fusion Development Agreement. The
views and opinions expressed herein do not necessarily reflect those
of the European Commission.  This work was also part-funded by the
Grant Agency of the Czech Republic under grant P205/11/2341, MSMT CR
{\#}7G10072 and the RCUK Energy Programme under grant EP/I501045.
\begin{table}
\begin{center}
\begin{tabular}{|l|l|l|l|l|}
\hline
Shot  & $I_\text{RMP}$ & parity & phase & note\\
\hline
22625 & 1.4 kA & even & $60^\circ$ & reversed $I_p$ and $B_T$ \\
\hline
25941 & 1.4 kA & even & $60^\circ$ & \\
\hline
25953 & 0 kA & N/A & N/A & \\
\hline
25965 & 1.4 kA & even & $0^\circ$ & \\
\hline
25056 & 0 kA & N/A & N/A & \\
\hline
25057 & 1.4 kA & odd & $60^\circ$ & \\
\hline
27652 & 1.4 kA & odd & $0^\circ$ & \\
\hline
\end{tabular}
\caption{List of shots used in this paper.}
\label{tab1}
\end{center}
\end{table}

\clearpage

\bibliography{/compass/home/cahyna/bibliografie/my.bib,/compass/home/cahyna/bibliografie/mypubs/mine.bib}
\bibliographystyle{pavel2}

\end{document}